\documentclass[usenatbib]{mnras}

\usepackage{natbib}
\bibpunct{(}{)}{;}{a}{}{,}
\usepackage{graphicx}
\usepackage{txfonts}
\usepackage{xcolor}
\usepackage{multirow}
\usepackage{pst-plot}
\usepackage{breakurl}
\SpecialCoor
\usepackage{hyperref}

\def\inte{{\em INTEGRAL}}

\def\swift{{\em Swift}}

\def\suzaku{{\em Suzaku}}
\def\nustar{{\em NuSTAR}}

\def\ferg{\mathrm{erg\,s^{-1}\,cm^{-2}}}

\def \src {IGR\,J16194-2810}

\definecolor{arancio}{rgb}{1,0.5,0}
\definecolor{viola}{rgb}{0.7,0,1}
\definecolor{verde}{rgb}{0.2,0.7,0.7}
\definecolor{cobalt}{rgb}{0.0, 0.28, 0.67}
\definecolor{airforceblue}{rgb}{0.36, 0.54, 0.66}
\definecolor{ballblue}{rgb}{0.13, 0.67, 0.8}
\definecolor{battleshipgrey}{rgb}{0.52, 0.52, 0.51}
\definecolor{darkgreen}{rgb}{0.0, 0.2, 0.13}

\title[X-ray observations of the SyXB \src]{\swift,\ \nustar,\ and \inte\ observations of the symbiotic X-ray binary \src}
\author[E. Bozzo et al.]{
E.\ Bozzo,$^{1}$\thanks{E-mail: enrico.bozzo@unige.ch}
P.\ Romano,$^{2}$
C.\ Ferrigno,$^{1,2}$
and L.\ Ducci,$^{1,3,2}$
\\
$^{1}$Department of Astronomy, University of Geneva, Chemin d'Ecogia 16, CH-1290 Versoix, Switzerland\\
$^{2}$INAF, Osservatorio Astronomico di Brera, Via E.\ Bianchi 46, I-23807, Merate, Italy\\
$^{3}$Institut f\"ur Astronomie und Astrophysik, Kepler Center for Astro and Particle Physics, University of Tuebingen, Sand 1, 72076 T\"ubingen, Germany
}

\date{}

\begin{document}

\maketitle

\begin{abstract}
We report on a simultaneous observational campaign with both \swift/XRT and \nustar\ targeting the symbiotic X-ray binary \src.\ The main goal of the campaign was to investigate the possible presence of cyclotron scattering absorption features in the broad-band spectrum of the source, and help advance our understanding of the process of neutron star formation via the accretion-induced collapse of a white dwarf. The 1-30~keV spectrum of the source, as measured during our campaign, did not reveal the presence of any statistically significant  absorption feature. The spectrum could be well described using a model comprising a thermal black-body hot component, most likely emerging from the surface of the accreting neutron star, and a power-law with no measurable cut-off energy (and affected by a modest absorption column density). Compared to previous analyses in the literature, we could rule out the presence of a colder thermal component emerging from an accretion disk, compatible with the idea that \src\ is a wind-fed binary (as most of the symbiotic X-ray binaries). Our results were strengthened by exploiting the archival XRT and \inte\ data, extending the validity of the spectral model used up to 0.3-40~keV and demonstrating that \src\ is unlikely to undergo 
significant spectral variability over time in the X-ray domain.
\end{abstract}


\section{Introduction}
\label{sec:intro}

\src\ is one of the rare symbiotic X-ray binaries (SyXBs), hosting a neutron star (NS) accreting from the wind of a red giant companion \citep[see, e.g.][and references therein]{yungelson19}. Although SyXBs are formally part of the low mass X-ray binaries (LMXBs) due to the presence in these systems of a low mass red giant, their behaviour in the X-ray domain strongly resembles that of wind-fed high mass X-ray binaries (HMXBs), specifically those hosting OB supergiants \citep[the so-called supergiant X-ray binaries, SgXBs; see][for recent reviews]{kre19}. SyXBs show long pulse periods, ranging from hundreds of seconds to hours, and display a pulsed fraction that can be as large as 30-50\,\%. They are characterized by peculiarly long orbital periods (several tens to thousands of days) and a prominent variability in the X-ray domain achieving a factor of $\sim$10-20 on timescales of thousands of seconds \citep[that is expected for a neutron star accreting from a highly structured stellar wind;][]{nunez17}. The similarity of SyXBs with other more commonly discovered classes of wind-fed binaries makes these objects difficult to be correctly identified. Over the past two decades, there have been reports of re-classifications of the donor stars in SyXBs (due to more and more accurate observational campaigns), and the list of such sources shrunk over time instead of being more populated \citep{masetti06,masetti07,masetti07b,nucita07,corbet08,marcu11,bozzo13,enoto14}. At the time of writing, only 5 objects have been firmly established as SyXBs \citep[see, e.g.][and references therein]{bozzo22}.

The X-ray spectra of SyXBs are reminiscent of those typically observed from other wind-fed binaries, being described by relatively hard models (extending up to several tens of keVs) with cut-off energies $\gtrsim$20~keV. The commonly derived high absorption column densities are  similar to those of other classes of wind-fed binaries, and ascribed to the presence of a dense stellar wind in the closest surroundings of the accreting compact objects \citep[see, e.g.][]{smith12,bozzo13b,bozzo22}. Due to these spectral characteristics and the similarity with the SgXBs, it has been proposed that SyXBs might host relatively young and strongly magnetized ($>$10$^{12}$~G) NSs. However, as extensively discussed by \citet{bozzo22}, this is difficult to reconcile with the long established fact that the NS magnetic field should decay with time (on a scale of 10$^6$~yr) and that the presence of a red giant requires a minimum
age of a SyXB of several Gyrs. A possibility is that the NS formed much later in the evolution of the binary due to the accretion induced collapse (AIC) of a white dwarf accreting from the stellar wind of the red giant \citep{nomoto91,fryer99}. The white dwarf can have an age matching that of the red giant and the NS would only form after a long-lasting accretion phase, thus preserving its high magnetic field before the system shines as a SyXB. The accretion induced collapse of a white dwarf into a NS is a poorly known process and it is not yet clear if a strongly magnetized NS can be produced by conservation of a sufficiently intense white dwarf magnetic flux during the collapse \citep[see, e.g.,][and references therein]{tauris15,aic1,aic2}.  Firmly establishing the presence of strongly magnetized NSs in SyXBs is thus a promising way to either challenge theories of the NS magnetic field decay or confirm the possibility that strongly magnetized
NSs can be formed via the AIC channel.

In order to investigate and possibly confirm the presence of strong magnetic field NSs in SyXBs, we initiated several observational campaigns targeting known objects in this class and performed a broad-band analyses of their X-ray emission to hunt for resonant scattering cyclotron lines \citep[CRSFs; see][for a recent review]{staubert19}. These features are known to provide the most direct evidence and robust estimate of NS magnetic fields in the range of few 10$^{12}$~G \citep[within an expected accuracy of about 30\%; see][]{poutanen13,mushtukov15}. So far, evidence for the presence of cyclotron lines in SyXBs has only been reported for IGR\,J17329-2731 \citep{bozzo18} and 4U\,1700+24 \citep{bozzo22}. The fundamental line in IGR\,J17329$-$2731 was found at $\sim$21~keV thus indicating a magnetic field as strong as 2.4$\times$10$^{12}$~G. In the case of 4U\,1700+24, the centroid energy of a tentative CRSF line was measured at $\sim$16~keV, translating into a possible magnetic field strength of 1.4$\times$10$^{12}$~G.

In this paper, we present the outcomes of our observational campaign carried out with \swift\ and \nustar\ in August 2023 and aimed at the SyXB \src. We provide a detailed analysis of the broad-band emission of the source with the main goal of looking for a possible presence of CRSFs. We also exploit all publicly available data from the $\sim$22~years long archive of \inte\ in order to further improve the energy coverage and carry out a thorough comparison with literature results on the source. A brief description of the target source is provided in Sect.~\ref{sec:src}, while an exhaustive overview of the data analysis and results is given in Sect.~\ref{sec:data}. Our conclusions are presented in Sect.~\ref{sec:conclusions}.

\section{\src}
\label{sec:src}

\src\ was discovered by \inte\ and first reported in the second IBIS/ISGRI catalogue published by \citet{bird06}. The source has since then been known to be a persistent and variable X-ray emitter, displaying a modest dynamic range in the X-ray luminosity by a factor $\lesssim$10 and a long-term average hard X-ray flux of $\sim$5$\times$10$^{-11}$~erg~cm$^2$~s$^{-1}$ in the 20-100~keV energy range \citep{bird16,bat105}. It was classified as a SyXB by \citet{masetti07}, following the availability of an accurate localization by the \swift\ satellite and the subsequent identification of the red giant companion at a distance of $\lesssim$3.7~kpc.  The system orbital period is not known yet, but recently \citet{luna23} reported the detection of a 242.837~min periodicity in the optical data of the source collected with the TESS mission that is likely associated with the NS spin period. \src\ would thus be hosting a very slow rotator, which is in agreement with the expectations for SyXBs \citep[see, e.g.][]{yungelson19}.

In the X-ray domain, observations of \src\ remains relatively sparse. Although it is routinely reported in both the \inte\ and \swift\ source catalogues, the source detection with coded masks instruments require exposures up to several Ms due to the low average flux in the hard X-ray domain. This challenged any attempt to perform detailed spectral analyses exploiting only these data.
The first broad-band analysis of the X-ray emission from \src\ was reported by \citet{masetti07}, combining two dedicated \swift/XRT pointings with the entire available \inte\ exposure time on the source collected at that time (from 2002 to 2006). The broad-band spectrum could be fit well with a single absorbed Comptonization model, although the constraints on the spectral energy distribution above 10~keV were limited due to the relatively low statistics of the \inte\ data and the energy interval 7-20~keV was not covered by this data-set.
\citet{masetti07} also reported about the source variability in the XRT data, reaching a factor of 8--10 in flux over a few thousands of seconds. This is typical (and expected) for wind-fed binaries. Similar results for both the spectral energy distribution and the short term variability were reported for a 46~ks \suzaku\ observation carried out in 2013 \citep{yuiko14}. These data allowed the authors to obtain more refined constraints on the parameters of the Comptonization model (several models were tested) and revealed the presence of a soft component dominating the source emission at energies $\lesssim$2~keV. This emission could be described by using a blackbody model which either mimicked the emission from a small portion of the NS surface (about 1.7~km) with a temperature around 1~keV or the presence of an accretion disk with a radius of $\lesssim$75~km and a peak temperature around 0.4~keV. The \suzaku\ data were still characterized by a relatively low statistics above 20~keV (due to limited sensitivity for faint objects of the HXD-PIN instrument), and a gap in the covered energy band between 9-20~keV. The search for CRSFs in the X-ray spectrum of \src\ has thus been hampered so far by the lack of a sensitive and uninterrupted coverage from the softest ($\lesssim$2~keV) to the hardest ($\gtrsim$10-20~keV) X-rays.

\section{Data processing and analysis}
\label{sec:data}

\subsection{\nustar}
\label{sec:nustar}

\src\ was observed by \nustar\ on 2023 August 08 from 03:01 to 19:21 UTC (ObsID~30901015002; PI: Bozzo).
We processed the data using standard methods and procedures\footnote{\url{https://heasarc.gsfc.nasa.gov/docs/nustar/analysis/nustar_swguide.pdf}}, exploiting the {\sc nupipeline} v.\,0.4.9 distributed with the {\sc Heasoft} software v.\,6.32.1.
The latest calibration files available at the moment of writing were used for the data reduction (caldb v.\,20230918).
After having applied all good time intervals (GTI) to the \nustar\ data  accounting for the Earth occultation and
the South Atlantic Anomaly passages, we obtained an effective exposure time of 30.0~ks.
The source photons were extracted from a 60\arcsec circle centred on the source,
while the background was evaluated using a 120\arcsec circle centered on a region in the same chip as the source but not contaminated from the source emission. No stray-light contamination was observed in either the FPMA or FPMB. Each module recorded during the observation an average source count-rate of 0.26$\pm$0.03~cts~s$^{-1}$ in the 2-30~keV energy band (all uncertainties in the paper are given at 90\% c.l., unless stated otherwise).

\begin{figure*}
  \hspace{-1.3cm}
  \includegraphics[width=9.7cm,angle=-90]{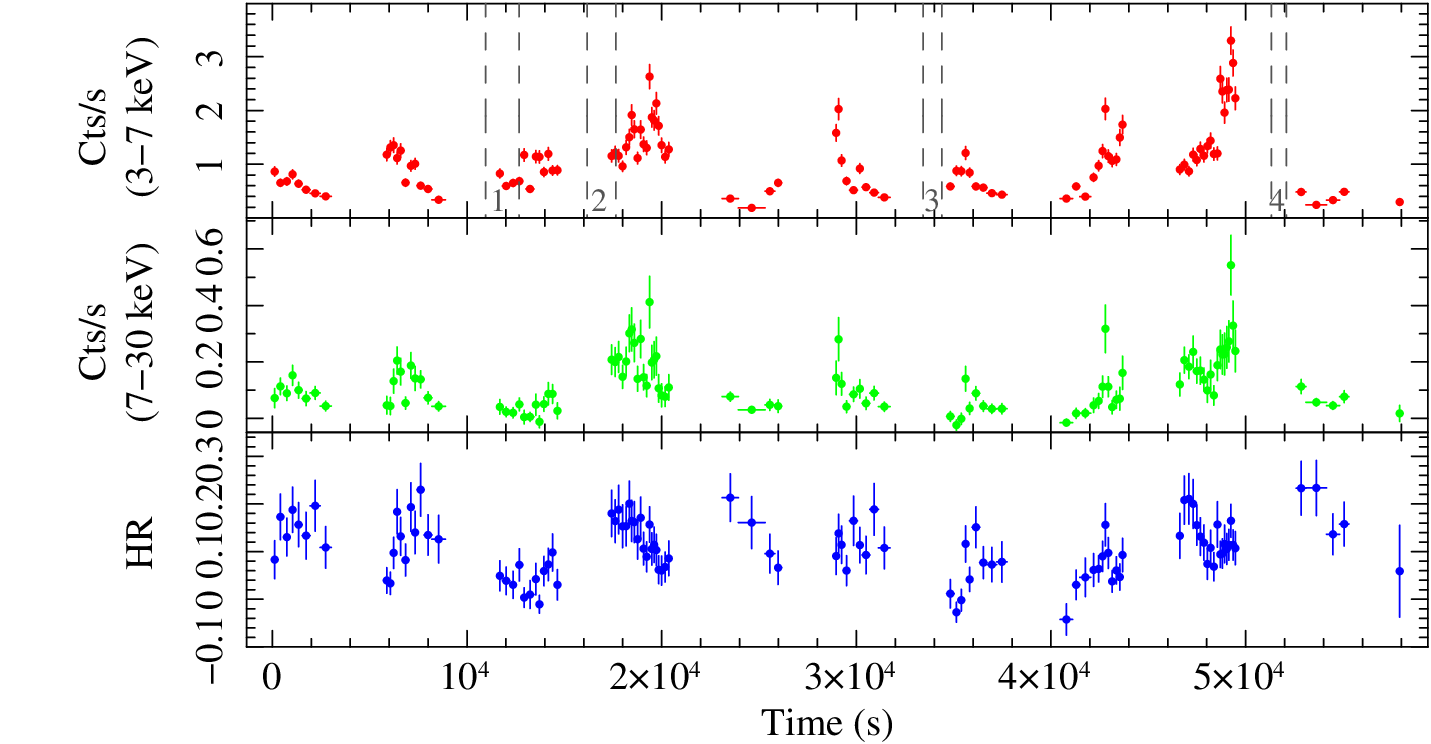}
  \caption{\label{fig:nustar_lc} Energy resolved lightcurves of \src\ as observed by \nustar during the ObsID~30901015002. The upper panel shows the lightcurve in the 3--10~keV energy range, while the mid-panel shows the lightcurve in the 10--30~keV energy range. Each of the displayed lightcurve is obtained by summing together data from the FPMA and FPMB, after having applied the correction for the background.} The bottom panel shows the HR. An adaptive rebinning has been used for all panels, set as to reach a minimum S/N=10 in the soft energy curve (see text for details). The dashed vertical lines mark the intervals over which the four orbits (labelled 1 to 4) of XRT data were collected.
\end{figure*}

We extracted the background corrected FPMA and FPMB energy-resolved lightcurves in the 3--10~keV and 10--30~keV bands, and summed together the lightcurves in the same energy band of the two FPMs to increase the statistics. We then used the summed energy-resolved lightcurves to compute the hardness-ratio (HR) via our developed adaptive rebinning algorithm presented in previous papers \citep[see, e.g.,][and references therein]{ferrigno22}. The lightcurves were rebinned by the algorithm to reach a signal-to-noise ratio (S/N) of at least 10 in the soft band curve and then the same rebinning is applied to the hard band curve before computing the HR. The result is reported in Fig.~\ref{fig:nustar_lc}. The source displays a variability similar to that reported previously in the literature, with variations up to a factor of 2--3 over timescales of few thousands of seconds. The variability of the HR is modest, and no major changes are observed across the \nustar\ observation that could justify a HR-resolved spectral analysis \citep[see, e.g.][]{bozzo17,ferrigno20}. We thus extracted a single spectrum for the FPMA and FPMB using the entire exposure time available during the observation. Modeling of these spectra is discussed in Sect.~\ref{sec:results}, by combining them with the available simultaneous \swift/XRT data (see Sect.~\ref{sec:swift}).  All spectra used in this paper were rebinned by using the optimal rebinning algorithm developed by \citet{Kaastra} and implemented within {\sc Heasoft} via the {\sc ftgrouppha} tool (using the option {\sc groupscale=25}). 

We also extracted the source and background event files from both the FPMA and FPMB, applying a barycentric correction with the tool {\sc barycorr}. These event files were inspected for the presence of possible periodicities in the range 10$^{-4}$-10\,Hz through the  Lomb-Scargle periodogram and epoch folding techniques. The periodogram of the source X-ray emission displayed a strong red noise dominating the frequencies up to 2$\times$10$^{-3}$~Hz. Above this frequency, no statistically significant ($\gtrsim$5$\sigma$) peak is observed. We note that the  tentative NS spin period of $\sim$243.837~minutes reported by \citet{luna23} could not be searched for in the \nustar\ data given the limited time coverage of ObsID~30901015002.

\subsection{\swift\ }
\label{sec:swift}

The \swift\ data of \src\ were processed and analysed using the standard software ({\sc FTOOLS}\footnote{\href{https://heasarc.gsfc.nasa.gov/ftools/ftools_menu.html}{https://heasarc.gsfc.nasa.gov/ftools/ftools\_menu.html.}} v6.31), calibration (CALDB\footnote{\href{https://heasarc.gsfc.nasa.gov/docs/heasarc/caldb/caldb_intro.html}{https://heasarc.gsfc.nasa.gov/docs/heasarc/caldb/caldb\_intro.html.}}  20230725), and methods. The \swift/XRT data were filtered with the task {\sc xrtpipeline} (v0.13.7). A log of all available XRT observations of \src{} is reported in Table~\ref{16194:tab:swift_xrt_log}.
%
%
%
%
%
 \begin{table*} 	
 \scriptsize
 \begin{center} 	
 \caption{{\it Swift}/XRT observation log: observing sequence, date (MJD of the middle of the observation), start and end times (UTC), and XRT exposure time. We also report in the table the results obtained by fitting the XRT spectrum in each observation with a model comprising an absorbed blackbody component (see text for details). Here, $N_{\rm}$ is the absorption column density, $kT$ is the blackbody temperature, and $R_{\rm BB}$ is the blackbody radius (evaluated for a distance of 3.7~kpc). The row ``Total'' reports the results for the XRT spectrum obtained by stacking together all other displayed observations. Note that the source was not detected by XRT during the ObsID~00036116006, so in this case a 3~$\sigma$ upper limit is reported on the flux by fixing the spectral parameters to the closest preceding observation in time.}  	
 \label{16194:tab:swift_xrt_log} 	
 \begin{tabular}{llllllllll} 
 \hline 
 \noalign{\smallskip} 
  Sequence   & MJD  & Start time  (UT)                    & End time   (UT)                 & Exposure  & $N_{\rm H}$ & $kT$ & $R_{\rm BB}$  &  Flux (0.3-10~keV) & C-statistics/d.o.f. \\ 
                   &           & (yyyy-mm-dd hh:mm:ss)  & (yyyy-mm-dd hh:mm:ss)  &(s)  & 10$^{22}$~cm$^{-2}$ & keV &  km &  10$^{-11}$$\ferg$ &          \\
  \noalign{\smallskip} 
 \hline 
 \noalign{\smallskip} 
00036116001	&	54129.45838	&	2007-01-29 03:36:10	&	2007-01-29 18:23:58	&	5140 & 0.11$^{+0.05}_{-0.05}$ & 1.21$^{+0.05}_{-0.05}$ & 0.58$^{+0.03}_{-0.04}$	& 5.6$^{+0.2}_{-0.3}$ & 606.1/613 \\
00036116002	&	54132.86922	&	2007-02-01 18:21:24	&	2007-02-01 23:21:57	&	2648 & 	0.2$^{+0.1}_{-0.1}$ & 1.14$^{+0.07}_{-0.07}$ & 0.55$^{+0.05}_{-0.05}$	& 3.4$^{+0.3}_{-0.3}$ & 395.0/427 \\
00036116004	&	59701.93159	&	2022-05-02 22:17:05	&	2022-05-02 22:25:52	&	527	& 0.3$^{+0.2}_{-0.3}$ & 1.0$^{+0.1}_{-0.1}$ & 0.82$^{+0.25}_{-0.20}$	& 4.5$^{+0.5}_{-1.1}$ & 97.7/151 \\
00036116005	&	59727.79012	&	2022-05-28 13:55:39	&	2022-05-28 23:59:52	&	925	& 0.2$^{+0.6}_{-0.2}$ & 0.8$^{+0.3}_{-0.2}$ & 0.29$^{+0.13}_{-0.06}$	& 0.25$^{+0.05}_{-0.25}$ & 37.4/41\\
00036116006	&	59756.73112	&	2022-06-26 13:20:44	&	2022-06-26 21:44:53	&	1344 & 	0.2 (fixed) & 0.8 (fixed) & --- & $<$0.03 \\
00036116007	&	59786.61003	&	2022-07-26 14:31:00	&	2022-07-26 14:45:53	&	893	& 0.3$^{+0.3}_{-0.3}$ & 0.9$^{+0.1}_{-0.1}$ & 0.57$^{+0.16}_{-0.10}$	& 1.8$^{+0.1}_{-0.5}$ & 136.8/187 \\
00089639003$^a$	&	60164.49869	&	2023-08-08 06:15:19	&	2023-08-08 17:40:53	&	4927 & 0.5$^{+0.2}_{-0.2}$ & 1.15$^{+0.08}_{-0.08}$ & 0.39$^{+0.04}_{-0.03}$	& 1.7$^{+0.1}_{-0.2}$ & 339.4/449 \\
  \noalign{\smallskip}
Total       &   & 2007-01-29 03:36:10 & 2023-08-08 17:40:53 & 16403 & 0.21$^{+0.05}_{-0.05}$ & 1.16$^{+0.03}_{-0.03}$ & 0.48$^{+0.02}_{-0.02}$	& 3.0$^{+0.1}_{-0.1}$ & 713.4/681 \\ 
  \noalign{\smallskip}
  \hline
  \end{tabular}
  \end{center}
  \begin{list}{}{} 
  \item[$^{\mathrm{a}}$] Simultaneous with the \nustar\ observation.
  \end{list} 
  \end{table*}

As shown in the Table, \src\ was observed only a few times with XRT. The two observations carried out in 2007 were already reported by \citet{masetti07}, while those carried out in 2022 remained so far still unpublished (at the best of our knowledge). The longest XRT exposure carried out in 2023 was obtained as part of our source monitoring campaign (obsID 00089639003; PI: Romano) and scheduled simultaneously with the \nustar\ data described in Sect.~\ref{sec:nustar}. We extracted for each observation the 0.3--10~keV source lightcurve (background subtracted) and the corresponding spectrum. The count-rate of the source recorded across all observations was relatively low ($\sim$0.1--0.5~cts~s${-1}$), and during the ObsID~00036116006 the source was not detected. The 3~$\sigma$ upper limit we derived by using the {\sc ximage} tool \citep{bozzo09} was of 6.6$\times$10$^{-3}$~cts~s$^{-1}$. For each observation where the source was detected, we performed a spectral fit within {\sc xspec} \citep[v.~12.13.1][]{xspec} using a simple model consisting of an absorbed blackbody ({\sc bbodyrad} in {\sc xspec}, see Sect.~\ref{sec:results}). Fits were carried out on the ungrouped spectra, using the C-statistics \citep{cash79}. For the absorption component, we used {\sc Tbabs} with the default {\em wilm} abundances \citep{wilms00} and {\em vern} cross sections \citep{vern96}. We report the results of this analysis in Table~\ref{16194:tab:swift_xrt_log}. For the ObsID~00036116006, we converted the upper limit on the source count-rate into an upper limit on the flux using the online tool {\sc webpimms} and assuming the same spectral parameters as in the ObsID~00036116005 (the closest preceding in time).

The source was significantly brighter during the 2007 observations, displaying an overall dynamic range of a factor of $\sim$180 in the X-ray luminosity and a relatively stable spectral energy distribution in the XRT energy band. Given the lack of any dramatic spectral variability across the XRT data, we also stacked together all observations to extract the highest possible S/N XRT spectrum. We fit this spectrum with the same simple model above and reported the results in Table~\ref{16194:tab:swift_xrt_log}.

\subsection{\inte\ }
\label{sec:inte}

In order to compare our results with those reported previously in the literature \citep{masetti07}, we exploited the full archive of the \inte\ data in order to extract the highest possible statistics spectra of \src\ from the JEM-X \citep{lund03} and IBIS/ISGRI \citep{ubertini03,lebrun03} instruments.  We analyzed all the publicly available \inte\ data starting from 2004 January 1st by using version 11.2 of the Off-line Scientific Analysis software (OSA) distributed by the ISDC \citep{courvoisier03} and implemented via the multi-messenger online data analysis system\footnote{\url{https://www.astro.unige.ch/mmoda/}.} \citep{mmoda}. \inte\ observations are divided into ``science windows'' (SCWs), i.e.,  pointings with typical durations of $\sim$2--3~ks. Only SCWs in which the source was located to within an off-axis angle of 3.5~deg from the center of the JEM-X field of view were included in the JEM-X analysis, while for IBIS/ISGRI we retained all SCWs where the source was within an off-axis angle of 12~deg from the center of the instrument field of view.  The source was detected at a significance of 9.8~$\sigma$ in the 25-40~keV IBIS/ISGRI mosaic and at a significance of 2.6~$\sigma$ in both the 3-20~keV JEM-X1 and JEM-X2 mosaics. An effective exposure of 3.26~Ms, 62.6~ks, and 64.9~ks was available for the source, in IBIS/ISGRI, JEM-X1, and JEM-X2, respectively\footnote{INTEGRAL products are available at the Legacy Gallery URL \url{https://www.astro.unige.ch/mmoda/gallery/astrophysical-entity/igr-j16194-2810}}.

The low significance of the detection in both IBIS/ISGRI and the two JEM-X demonstrates that \src\ is a very faint source for \inte.\ The statistics of the data was far too low to perform any meaningful study of intensity and spectral variability in the \inte\ data, as well as the search for coherent modulations (e.g., associated to the possible 244~min spin period). We thus extracted a single spectrum for the three instruments using the entire exposure time available.  These spectra are used in Sect.~\ref{sec:results} to carry out the analysis of the broad-band spectral energy distribution of the source.

\subsection{Results}
\label{sec:results}

In order to determine the best model to describe the broad-band continuum from \src\ and eventually look for the possible presence of CRSFs, we performed a combined fit of the \nustar\ data with the simultaneous XRT observation (ObsID~00089639003, see Table~\ref{16194:tab:swift_xrt_log}).
\begin{figure}
  \centering
  \includegraphics[width=5.6cm,angle=-90]{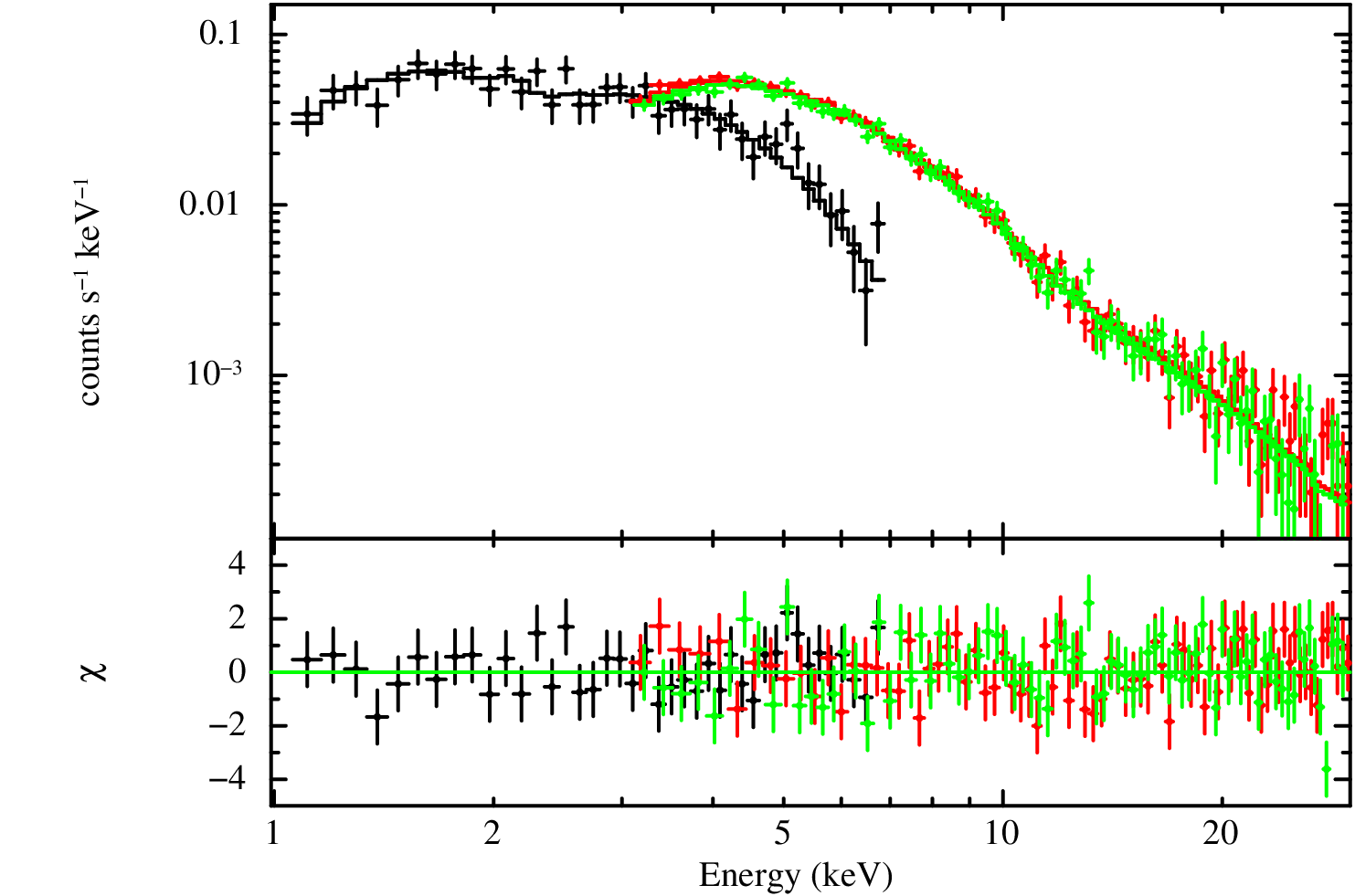}
  \caption{\label{fig:bband1} Broad-band X-ray spectrum of \src\ as observed simultaneously by \swift/XRT (black), \nustar/FPMA (red) and \nustar/FPMB (green). The best fit model is an absorbed hot blackbody plus a power law (see text for details). The residuals from the fit are shown in the bottom panel.}
\end{figure}

We first attempted to describe the spectrum with a simple power-law model, but this yielded a largely unacceptable result ($\chi^2$/d.o.f.=383.0/204, where d.o.f. are the degrees of freedom). Including the presence of a high energy cut-off ({\sc cutoffpl} in {\sc xspec}) did not significantly improve the results ($\chi^2$/d.o.f.=391.7/196), leaving s-shaped residuals all along the covered energy band. Progressing with the usage of phenomenological models, we tested that adding a thermal component at low energies largely improved the fit ($\chi^2$/d.o.f.=205.9/194). A hot blackbody plus a power-law component provided an acceptable fit to the data, assuming the blackbody is coming from a relatively small region (a fraction of a km) and characterized by a temperature of $\simeq$1.08~keV (see Table~\ref{tab:spe_summary} and Fig.~\ref{fig:bband1}). In this fit, no cut-off energy was required (the fit converged to the highest boundary of 500~keV for the cut-off energy). Exchanging the hot blackbody with a disk-blackbody ({\sc diskbb}) component did not provide an equivalently good fit, meaning that we can exclude the thermal excess at low energy is emerging from an accretion disk surrounding the compact object \citep[a model that could satisfactorily fit the \suzaku\ data reported by][]{yuiko14}. By inspecting the residuals from our best fit in Fig.~\ref{fig:bband1}, we noticed some structures remaining around $\sim$12~keV that could mimic the presence of an absorption feature. We thus tested the inclusion of an additional {\sc gabs} component in the fit, which is usually adopted in the literature to take into account the presence of CRSFs in the X-ray spectra on highly magnetized NS \citep[see, e.g.,][and references therein]{staubert19}. The fit including the hypothetical absorption feature returned a centroid energy of 11.1$^{+0.3}_{-0.4}$~keV and a practically unconstrained depth (width) of 0.5$^{+0.4}_{-0.5}$~keV (0.03$^{+0.9}_{-0.03}$~keV). The addition of this feature provided only a marginal change to the quality of the fit ($\chi^2$/d.o.f.=200.0/192) and the derived feature parameters were poorly convincing as it turned out to be significantly less wide and deep that CRSFs commonly observed in other similar wind-fed sources \citep[see, e.g.,][and references therein]{ferrigno22}. It was also checked that different choices for the extraction regions of the source and background do not significantly alter the results, i.e.\ residuals around 12~keV remained somewhat visible in all cases but never statistically significant for the addition of a {\sc gabs} component. We concluded that no convincing evidence of CRSFs could be found in the broad-band spectrum of the source as measured simultaneously by \swift/XRT and the two \nustar\ FPMs. We included in all fits normalization constants in order to take into account the inter-calibrations between \nustar\ and \swift,\ as well as the fact that observations are simultaneous but do not cover exactly the same time due to the fragmentation of the XRT observation and the Earth occultation along the orbits of both satellites. As shown in Table~\ref{tab:spe_summary}, these constants turned out to be fully compatible with unity.

\begin{table}
    \begin{center}
    \caption{\label{tab:spe_summary} Spectral fit results obtained for the simultaneous \swift/XRT and \nustar\ observations carried out in August 2023 (``2023 campaign'') and for the combined fit with all data available (including \inte\ and the stacked spectrum summing up all available XRT data). In both cases, the best fit is obtained with a model comprising an absorbed blackbody and a power-law component (see text for details). In addition to the spectral parameters already introduced in Table~\ref{16194:tab:swift_xrt_log}, here we also indicated the normalization constants ($C$) that are referred to that of the \nustar/FPMA (fixed to one in all fits), as well as the power-law photon index and normalization ($\Gamma$, $N_{\Gamma}$). The blackbody radius is obtained for a distance of 3.7~kpc. The flux is given in the 3-20~keV energy band, not corrected for absorption.}
    \begin{tabular}{lr@{}lr@{}ll}
        \hline
        \hline
        Parameter & \multicolumn{2}{c}{Best fit values} & \multicolumn{2}{c}{Best fit values} & Units \\
                        & \multicolumn{2}{c}{{\sc 2023 campaign}} & \multicolumn{2}{c}{{\sc Total}} &  \\
        \hline
        $N_{\rm H}$ & 1.0 &$_{-0.3}^{+0.3}$ & 0.6 &$_{-0.2}^{+0.2}$ & 10$^{22}$~cm$^{-2}$ \\
        $kT$ & 1.08 &$_{-0.03}^{+0.03}$ &  1.10 &$_{-0.02}^{+0.02}$ & keV \\
        $R_{\rm BB}$ & 0.35 &$_{-0.02}^{+0.02}$ & 0.33 &$_{-0.01}^{+0.01}$ & km \\
        $\Gamma$ & 1.9 &$_{-0.2}^{+0.2}$ & 1.8 &$_{-0.2}^{+0.2}$ & \\
        $N_{\rm pow}$ & 1.8 &$_{-0.7}^{+1.1}$ &  1.4 &$_{-0.5}^{+0.7}$ & 10$^{-3}$ \\
        $C_{\rm FPMB}$ & 1.00 &$_{-0.03}^{+0.03}$ & 1.00 &$_{-0.03}^{+0.03}$ & \\
        $C_{\rm XRT}$ & 1.03 &$_{-0.08}^{+0.08}$ & 1.76 &$_{-0.07}^{+0.08}$ &  \\
        $C_{\rm ISGRI}$ & --- & & 4.7 &$_{-0.9}^{+1.0}$  &  \\
        $C_{\rm JEM-X1}$ & --- & & 2.5 &$_{-1.0}^{+1.0}$ &  \\
        $C_{\rm JEM-X2}$ & --- & & 2.1 &$_{-0.8}^{+0.8}$  &  \\
        Flux & 1.5 & $_{-0.2}^{+0.1}$ & 1.5 & $_{-0.1}^{+0.1}$ & 10$^{-11}$~$\ferg$ \\
        $\chi^2$/d.o.f. & 205.9/ & 195 & 261.3/& 225 & \\
        \hline
        \hline
    \end{tabular}
\end{center}
\end{table}
\begin{figure*}
\centering
  \includegraphics[width=10cm,angle=-90]{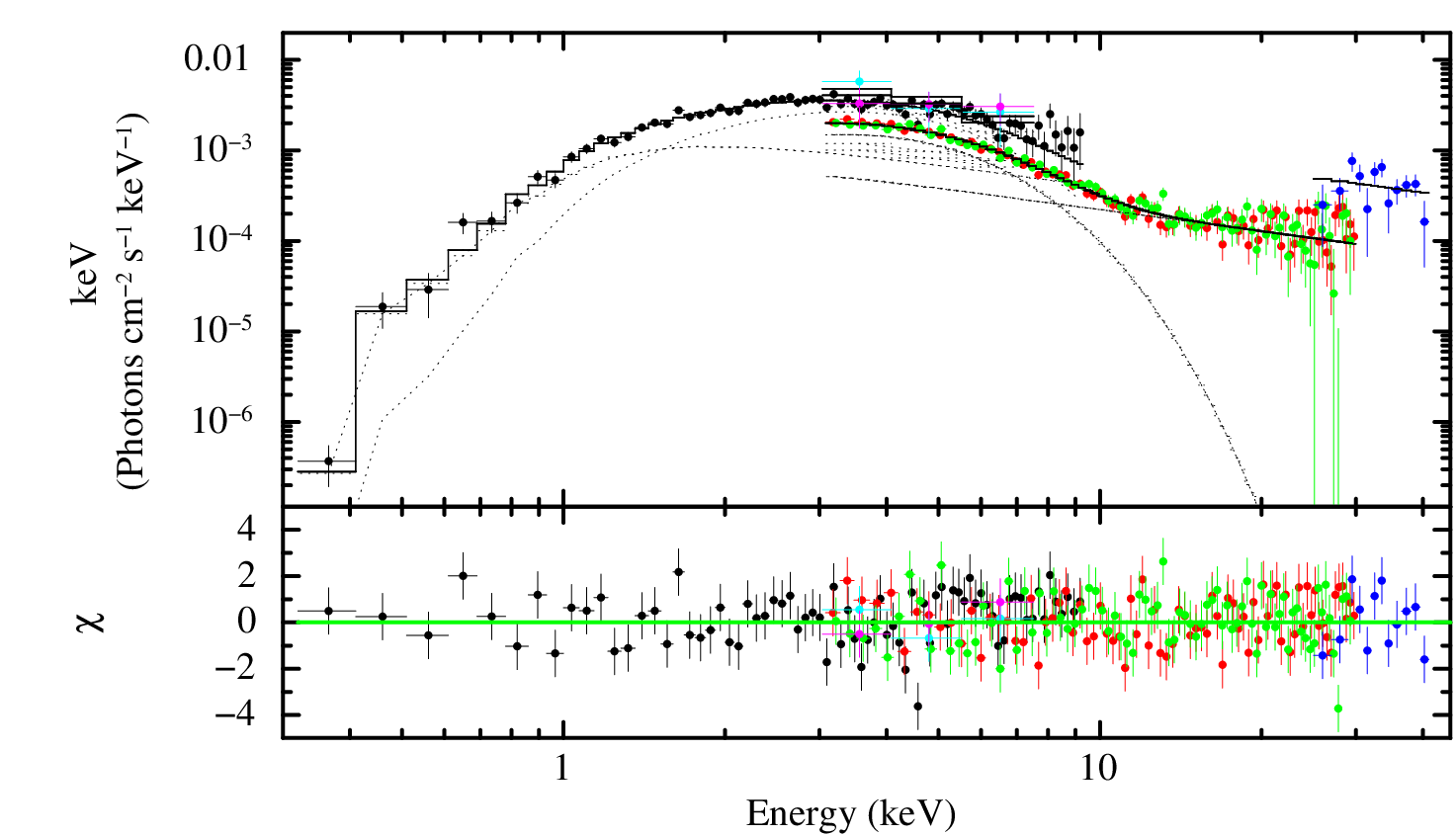}
  \caption{\label{fig:bband2} Same as Fig.~\ref{fig:bband1}, but showing the unfolded spectra and including now the XRT spectrum integrated over all data available (the ``total'' spectrum in Table~\ref{tab:spe_summary}). The \inte\ data are included as well (IBIS/ISGRI in blue, JEM-X1 in magenta, and JEM-X2 in cyan). The best fit model is an absorbed hot blackbody plus a power law (see text for details). The residuals from the fit are shown in the bottom panel.}
\end{figure*}

Following previous results in the literature (see Sect.~\ref{sec:src}), we tested the suitability of more physically-motivated spectral models. We used a two components model, comprising the hot blackbody identified above and a Comptonization component, testing both the addition of Comptonization of soft seed photons in a hot plasma \citep[{\sc comptt} in  {\sc xspec},][]{tita94apj} and the effect of a thermally Comptonized continuum \citep[{\sc nthcomp} in  {\sc xspec},][]{nthcomp1,nthcomp2}. In both cases the temperature of the seed photons for the Comptonization was linked to the temperature of the hot blackbody. From these fits we found that for both the Comptonization components it was not possible to constrain the electron temperature of the plasma and the fits always converged to a solution with the smallest allowed plasma optical depth and the highest allowed electron temperature. Given the results of the phenomenological models illustrated before, we concluded that this is due to the lack of a clear cut-off energy in the broad-band data and we thus do not discuss the Comptonization models further.

Motivated by the lack of any dramatic variability in the spectral energy distribution of the source from the XRT data, we also tested the suitability of the hot blackbody plus power-law model to the combination of the XRT spectrum extracted by stacking all data available (the ``total'' spectrum in Table~\ref{16194:tab:swift_xrt_log}) with the \nustar\ spectra and the spectra derived from the \inte\ long-term archive. The same phenomenological model exploited for our 2023 campaign could also satisfactorily fit the broader-band spectrum using all data available. We report the results of this analysis in Fig.~\ref{fig:bband2} and in Table~\ref{tab:spe_summary}. Also for these combined spectra, no statistically significant  evidence of a CRSF in the X-ray spectrum of \src\ could be found. Note that the somewhat large normalization constants derived for IBIS/ISGRI and JEM-X are expected, as these instruments are likely to have detected the source across the years during the more intense emission periods (due to the reduced sensitivity compared, e.g., to \swift/XRT and the \nustar/FPMs).

\section{Conclusions}
\label{sec:conclusions}

In this paper, we reported on the results of our observational campaign on the SyXB \src, carried out by exploiting the \swift\ and \nustar\ capabilities with the main goal of investigating with an unprecedented accuracy the broad-band X-ray spectral energy distribution of the source and search for possible CRSFs. As illustrated in Sect.~\ref{sec:intro}, confirming the presence of highly magnetized NSs in SyXBs holds the potential of improving our understanding on the NS formation via the AIC channel.

The collected \swift/XRT and \nustar/FPMs data in 2023 found the source 
at an average flux about a factor of $\sim$3 lower than previously observed in 2007. A variability at the same level is apparent also within the \nustar\ data. Overall, the archival XRT data demonstrated a total dynamic range in the X-ray luminosity by a factor of $\simeq$180, a value roughly an order of magnitude higher than the one reported so far in the literature \citep{masetti07} but still well compatible with what is usually observed in wind-fed binaries \citep[see, e.g.][and references therein]{kre19}. The long-term variability of the source could not be studied with \inte\ due to the faintness of the source for all instruments on-board this satellite.

The spectral analysis of the simultaneous XRT and FPMs data revealed that the broad-band X-ray spectrum of the source could be described well by using a hot absorbed blackbody model, completed by a power law at the higher energies ($\gtrsim$2~keV). The temperature ($\gtrsim$1~keV) and radius ($\ll$1~km) of the blackbody component are typical of those expected for the X-ray radiation emerging from the surface of an accreting NS \citep[see also the discussion in][]{yuiko14}. No convincing residuals from the fit could be identified to justify the inclusion of an absorption component that could mimic the presence of a CRSF. Contrary to previous findings in the literature, our analysis could rule out the presence of a low temperature soft component that could have emerged from an accretion disk. This is likely due to the fact that previous analyses exploited data which did not extend below 0.9~keV and could not take advantage of a full uninterrupted energy coverage from 1~keV up to 30~keV \citep{masetti07,yuiko14}. For the same reason, we were unable to obtain acceptable fits by using a single Comptonization model, as done in \citet{masetti07}, or exploiting the combination of a thermal blackbody and a Comptonization component as done in \citet{yuiko14}. All physical Comptonization models we tested could not be satisfactorily constrained due to the fact that no cut-off energy is measured in the energy range covered by \nustar.\  This did not allow us to obtain meaningful values for the plasma optical depth and the electron temperature, both converging toward the lowest and highest boundaries allowed by the models, respectively.

The above results were further confirmed by extending the energy coverage and the statistics of the broad-band spectra of \src\ thanks to the availability of the \inte\ data covering the period 2004-2023 and making usage of the stacked spectrum obtained from all available XRT data from 2007 up to 2023. The same phenomenological model used to describe the simultaneous XRT+FPMs data could also successfully describe the longer-term observation, proving that the source is unlikely to undergo over time major changes in its X-ray spectral energy distributions.

As of today, only two SyXBs have shown evidence of CRSFs but a few sources in this class still require high statistics broad-band observations to carry out such investigations. Additional \nustar\ data could help carrying out this study in the near future.

\section*{Data availability}
All data exploited in this paper are publicly available from the \nustar,\ \swift,\ and \inte,\ archives and processed with publicly available software.

\section*{Acknowledgements}
We thank the anonymous referee for useful comments.
EB and PR acknowledges financial contribution from contract ASI-INAF I/037/12/0.
This work made use of data supplied by the UK Swift Science Data Centre at the
University of Leicester \citep[see][]{2007A&A...469..379E,2009MNRAS.397.1177E}.

\bibliography{bib}{}
\bibliographystyle{mnras}

\appendix

\end{document}